\documentclass[onecolumn,preprintnumbers,preprint,groupaddress,11pt, nofootinbib]{revtex4-1}

\usepackage{amsmath}
\usepackage{graphics}
\usepackage{graphicx}
\usepackage[mathcal]{eucal}
\usepackage{color}
\usepackage{appendix}
\usepackage{footmisc}

\begin{document}
\preprint{UCI-TR-2010-25}

\title{Lifting Slepton Masses with a Non-universal, Non-anomalous $U(1)^{\prime}_{\mbox{\tiny NAF}}$ \\
in Anomaly Mediated SUSY breaking} 

\author{Mu-Chun Chen}
\email[]{muchunc@uci.edu}
\author{Jinrui Huang}
\email[]{jinruih@uci.edu}
\affiliation{Department of Physics and Astronomy, University of California, Irvine, CA 92697-4575, U.S.A.}

\date{\today}

\begin{abstract}
\label{abstract}
We extend the Minimum Supersymmetry Standard Model by a non-anomalous family (NAF) $U(1)^{\prime}_{\mbox{\tiny NAF}}$ gauge symmetry. All gauge anomalies are cancelled with no additional exotics other than the three right-handed neutrinos. The FI D-terms associated with the $U(1)^{\prime}_{\mbox{\tiny NAF}}$ symmetry lead to additional positive contributions to slepton squared masses. In a RG invariant way, this thus solves the tachyonic slepton mass problem in Anomaly Mediated Supersymmetry Breaking. In addition, the $U(1)^{\prime}_{\mbox{\tiny NAF}}$ symmetry naturally gives rise to the fermion mass hierarchy  and mixing angles, and  determines the mass spectrum of the sparticles.
\end{abstract}


\maketitle

\section{Introduction}
\label{sec:Intro}
Supersymmetry (SUSY) is one of the most appealing candidates as the new physics beyond the standard model (SM). As no sparticle has been discovered at energy scales accessible to the current collider experiments, SUSY must be broken at low energy. There are several mechanisms for mediating SUSY breaking that have been proposed. Among these mediation mechanisms, Anomaly Mediated SUSY Breaking (AMSB)~\cite{ref:SUSYAMSB} turns out to be an extremely predictive framework, in which the soft masses for the sparticles are generated by the conformal anomaly. As a result, {\it all} soft masses are determined entirely by the low energy dynamics ({\it i.e.} that of the MSSM) and one single parameter, $M_{\mbox{\tiny aux}}$, the F-term of some compensator chiral superfield. This is in stark contrast to the generic MSSM, where 124 parameters are present mostly to account for the soft SUSY breaking sector. 

The high predictivity also leads to a severe problem in AMSB models as generically the slepton masses are predicted to be tachyonic, because the electroweak gauge groups, $SU(2)_{L}$ and $U(1)_{Y}$, of the MSSM are not asymptotically free. Squarks do not suffer from the same problem as $SU(3)_{c}$ is asymptotically free. To solve the slepton mass problem, varieties of  approaches have been proposed ~\cite{ref:soluAMSB}. 
For example, the simpliest case is by adding an arbitrary universal scalar mass squared term to all sfermion masses. Nevertheless, the UV insensitivity in the predictions for  the soft masses is lost in this scenario. Additional positive contributions to slepton squared masses can also arise by introducing new particles at the TeV scale with large Yukawa couplings to the lepton chiral superfields~\cite{ref:YukawaAMSB} or by imposing an asymptotically free horizontal gauge symmetry based on $SU(2)_H$ or $SU(3)_H$~\cite{ref:SUHorizontal}. 

An extra $U(1)^{\prime}$ symmetry has been proposed before as a renormalization group (RG) invariant solution  to the slepton mass problem, with the Fayet-Illiopoulos (FI) D-terms ~\cite{ref:FIDterm} associated with the $U(1)^{\prime}$ symmetry rendering all slepton squared masses positive. In the previous works, the extra $U(1)$ symmetry considered is generation independent (and thus it is a linear combination of $U(1)_{Y}$ and $U(1)^{\prime}$ such as $U(1)_{B-L}$)~\cite{ref:U1U1B-L2,ref:U1U1B-L3}. A generation dependent extra $U(1)$ has also been utilized~\cite{ref:U1Gen}; nevertheless, earlier works only consider anomalous $U(1)$, where only the mixed anomalies are cancelled by the Green-Schwarz mechanism and additional exotic fields in addition to the RH neutrinos must be present to cancel the $[U(1)^{\prime}]^{3}$ anomaly. 

In this note, we introduce a non-anomalous family (NAF) symmetry $U(1)^{\prime}_{\mbox{\tiny NAF}}$ in the presence of three RH neutrino chiral superfields. In addition to solving the slepton mass problem, the $U(1)^{\prime}_{\mbox{\tiny NAF}}$ symmetry plays the role of a family symmetry naturally giving rise to fermion masses and mixing angles through the Froggatt-Nielsen mechanism~\cite{ref:frogNiel}. The anomaly cancellation conditions give rise to constraints on the $U(1)^{\prime}_{\mbox{\tiny NAF}}$ charges of the chiral superfields, more stringent than in the case of an anomalous $U(1)^{\prime}$.  While there exists an earlier claim~\cite{Ibanez:1994ig} that the $U(1)$ symmetry has to be anomalous in order to generate realistic fermion masses and mixing, we note that counter examples to this claim have been found in Ref.~\cite{ref:gauTrmNeuM, ref:SU5U1, ref:TeVU1} in which it is shown that a non-anomalous $U(1)$ symmetry can be a family symmetry giving rise to realistic masses and mixing angles of the SM fermions. Given that the $U(1)^{\prime}_{\mbox{\tiny NAF}}$ breaking scale in our model is close to the GUT scale, flavor violation mediated by the $Z^{\prime}$ gauge boson associated with the non-universal $U(1)^{\prime}_{\mbox{\tiny NAF}}$ is highly suppressed. 

The paper is organized as follows. In Sec.~\ref{sec:NegMass}, we review the generic features of AMSB and the solution to the problem of the negative slepton squared masses with an additional $U(1)^{\prime}_{\mbox{\tiny NAF}}$ symmetry. We introduce our model based on a non-universal, non-anomalous $U(1)^{\prime}_{\mbox{\tiny NAF}}$ symmetry in Sec.~\ref{sec:U1prime}, which is followed by Sec.~\ref{sec:FermMassMixing} where the predictions of fermion mass hierarchy and mixing angles are given. We present our numerical results for the sparticle spectrum in Sec.~\ref{sec:SparMass}. Finally, Sec.~\ref{sec:Conclusion} concludes the paper.

\section{Slepton Squared Masses in Anomaly Mediated SUSY Breaking}
\label{sec:NegMass}

The general soft SUSY breaking Lagrangian is given by,
\begin{equation}
\mathcal{L}_{soft} = -(m^2)_j^i \phi^i \phi^j - \biggl( \frac{1}{2} b^{ij} \phi^{i} \phi^{j} + \frac{1}{6} h^{ijk}\phi_i \phi_j \phi_k + \frac{1}{2} M_a \lambda_a \lambda_a + h.c. \biggr) \; ,
\end{equation}
where $M_a$ $(a=1,2,3)$ are the mass terms of the gaugino $\lambda_{a}$, $b^{ij}$ and $h^{ijk}$ are the bi-linear and  tri-linear terms, respectively, and $(m^2)^{i}_{j}$ are the scalar squared mass terms. One of the salient features of AMSB is that it predicts the following relations for the soft breaking terms which are renormalization group (RG) invariant~\cite{ref:U1U1B-L1, ref:RGEAMSB}, 
\begin{eqnarray}
\label{eqn:RGE1}
M_a = m_{3/2} \beta_{g_a}/g_a, \\
\label{eqn:RGE2}
h^{ijk} = -m_{3/2} \beta_Y^{ijk}, \\
\label{eqn:RGE3}
(m^2)_j^i = \frac{1}{2} m_{3/2}^2 \mu \frac{d}{d \mu} \gamma_j^i, \\
\label{eqn:RGE4}
b^{ij} = \kappa m_{3/2} \mu^{ij} - m_{3/2} \beta_{\mu}^{ij},
\end{eqnarray}
where $\gamma_j^i$ are the anomalous dimensions of the chiral superfields, $\mu^{ij}$ are the $\mu$ terms and $\beta_{g_a}$, $\beta_{Y}$ are the $\beta$-functions of the gauge and Yukawa couplings, respectively, and $\beta_Y$ is given by 
\begin{equation}
\beta_Y^{ijk} = \gamma_l^i Y^{ljk} + \gamma_l^j Y^{ilk} + \gamma_l^k Y^{ijl},
\end{equation}
and $\beta_{\mu}$ has a similar expression. With proper normalization, the F-term $M_{\mbox{\tiny aux}}$ is taken to be the gravitino mass, $m_{3/2}$. 

In the presence of the $U(1)^{\prime}_{\mbox{\tiny NAF}}$, there are additional Fayet-Illiopolous (FI) D-term contributions to the scalar squared masses.
Including the additional FI-D term contributions to the scalar masses, the new scalar squared masses at the GUT scale can be written as ~\cite{ref:U1U1B-L1, ref:RGEAMSB} 
\begin{eqnarray}
\bar{m}_{Q}^2 & = & m_{Q}^2 + \zeta q_{Q_i} \delta_{j}^{i} \; , \nonumber \\
\bar{m}_{u^c}^2 & = & m_{u^c}^2 + \zeta q_{u_i} \delta_{j}^{i} \; , \nonumber \\
\bar{m}_{d^c}^2 & = & m_{d^c}^2 + \zeta q_{d_i} \delta_{j}^{i} \; , \nonumber \\
\bar{m}_{L}^2 & = & m_{L}^2 + \zeta q_{L_i} \delta_{j}^{i} \; , \nonumber \\
\bar{m}_{e^c}^2 & = & m_{e^c}^2 + \zeta q_{e_i} \delta_{j}^{i} \; ,  \nonumber \\
\bar{m}_{H_u}^2 & = & m_{H_u}^2 + \zeta q_{H_u} \; , \nonumber \\
\bar{m}_{H_d}^2 & = & m_{H_d}^2 + \zeta q_{H_d} \; . 
\label{eqn:FIDMass}
\end{eqnarray}
where $q_{Q_{i}}$, $q_{u_{i}}$, $q_{d_{i}}$, $q_{L_{i}}$, $q_{e_{i}}$, and $q_{N_{i}}$ denote, respectively, the charges of the quark doublet $(Q_i)$, iso-singlet up-type quark $(u_i^c)$, iso-singlet down-type quark $(d_i^c)$, lepton doublet $(L_i)$, iso-singlet charged lepton $(e_i^c)$, and right-handed neutrino $(\nu_i^c)$. $m_{Q}^2$, $m_{u^c}^2$, etc denote the AMSB contributions to the scalar squared masses. With the additional $U(1)^{\prime}$ D-term contribution, the RG invariance is still preserved. The parameter $\zeta$ is the effective Fayet-Iliopoulos term setting the overall scale of the D-term contribution and it is a free parameter. 

For reasonable $U(1)^{\prime}_{\mbox{\tiny NAF}}$ D-term contribution assumption, we can solve the tachyonic slepton mass problem. In our analysis, the value of the effectve D-term, $\zeta$, which is a field dependent quantity, is on the order of the $M_{SUSY}$ scale. One of the mechanisms to naturally realize this is shown below ~\cite{ref:U1U1B-L2,ref:U1U1B-L3,ref:dimTrans}. The $U(1)_{\mbox{\tiny NAF}}^{\prime}$ breaking is achieved through the following terms in the superpotential,
\begin{equation}
W = S(\Phi \Phi^{\prime} - \Lambda^{2}) \; ,
\end{equation}
where $S$ is a gauge singlet. In the supersymmetric limit, 
\begin{equation}
\left< \phi \right> = \left< \phi^{\prime} \right> = \Lambda \sim \mathcal{O}(M_{GUT}) \; .
\end{equation} 
Naively, if the soft masses of the $\phi$ and $\phi^{\prime}$ fields, $m_{\phi}^{2}$ and $m_{\phi^{\prime}}^{2}$, are different, these VEVs then get shifted by different amounts, and the resulting effective D-term contribution is, 
\begin{equation} 
\label{eqn:zeta}
\zeta \sim (\left< \phi \right>^2 - \left< \phi^{\prime} \right>^2 ) \sim (m_{\phi}^{2} - m_{\phi^{\prime}}^{2}) \sim \mathcal{O}(M_{SUSY}) \; .
\end{equation} 
However, if the $\phi$ and $\phi^{\prime}$ fields are very heavy, say, on the order of  $\sim \mathcal{O}(M_{GUT})$, the $\phi$ and $\phi^{\prime}$ fields will decouple below the GUT scale, and there is no D-term contribution to the scalar masses. One way to have a non-zero D-term is through the deflection~\cite{ref:deflc}. In general, this modifies the AMSB trajectory unless $r \, \ll \, 1$ where $F_{S}/\left<S\right> = (1 + r) m_{3/2}$ (recall that $m_{3/2}$ is the F-term of the compensator field) so that the correction to the AMSB trajectory can be neglected. In our model, we take this approach to generate a non-zero effective D-term. While strictly speaking the predictions for the sfermion masses are UV sensitive as the size of the effective D-term depends on the UV physics, the correction to the AMSB trajectory is negligible and the predictions for the sfermion masses are still RG invariant with the additional D-term contributions, because the $U(1)^{\prime}$ symmetry is anomaly free~\cite{ref:U1U1B-L1,ref:RGEAMSB}. In our model, the difference between $m_{\phi}^2$ and $m_{\phi^{\prime}}^2$ is on the order of SUSY breaking which is consistent with Eq. (\ref{eqn:zeta}). We note that
\begin{equation}
\label{eqn:zetaNew}
\zeta \propto \frac{r^2}{4g} \left(\frac{m_{3/2}}{16 \pi^2}\right)^2 \sim \frac{r^2}{4g} \left(\frac{40 \; \mbox{TeV}}{16 \pi^2}\right)^2 \, ,
\end{equation}
and consequently, to have $\zeta \sim \mathcal{O}(M_{SUSY})$ so as to to solve the tachyonic slepton mass problem while preserving AMSB trajectory ($r \, \ll \, 1$), we have to choose $g \ll \mathcal{O}(0.1)$ where $g$ is the $U(1)^{\prime}$ gauge coupling. This is realized in our model without any fine-tuning as demonstrated below.


\section{The Non-anomalous $U(1)^{\prime}_{\mbox{\tiny NAF}}$ Model}
\label{sec:U1prime}

In the presence of the $U(1)^{\prime}_{\mbox{\tiny NAF}}$ symmetry, the superpotential that gives masses to all fermions and Higgses is given as follows,
\begin{eqnarray}
\label{eqn:sPoten}
W =  Y_uH_uQu^c + Y_dH_dQd^c + Y_eH_dLe^c + Y_{\nu}H_uL\nu^{c} + Y_{N} \Psi \nu^c\nu^c + \mu H_uH_d \; + \mu^{\prime} \Phi \Phi^{\prime} \; . \; 
\end{eqnarray}
Note that in the above equation, the family indices are suppressed. All chiral superfields including the additional three right-handed neutrinos, $\nu^c$, as well as the flavon fields, $\Phi$, $\Phi^{\prime}$, and $\Psi$,  are charged under the $U(1)^{\prime}_{\mbox{\tiny NAF}}$ symmetry. We assume that all flavon fields, $\Phi$, $\Phi^{\prime}$, $\Psi$, and the Higgs fields, $H_u$ and $H_d$, appear in conjugate pairs, that is, they all have one partner carrying opposite $U(1)_{\mbox{\tiny NAF}}^{\prime}$ charge correspondingly. Consequently, their fermionic components do not contribute to the gauge anomalies. Here we consider generation dependent $U(1)^{\prime}_{\mbox{\tiny NAF}}$ so that the $U(1)^{\prime}_{\mbox{\tiny NAF}}$ symmetry also plays the role of a family symmetry (see the next section). 

There are in total six anomaly cancellation conditions ~\cite{ref:TeVU1}: 
\begin{eqnarray}
\label{eqn:su3u1}
[SU(3)]^{2} U(1)^{\prime}_{\mbox{\tiny NAF}} & : &   \sum_{i} [ 2q_{Q_i} - (-q_{u_i}) - (-q_{d_i}) ]  = 0 \; , 
\\
\label{eqn:su2u1}
[SU(2)_{L}]^{2} U(1)^{\prime}_{\mbox{\tiny NAF}} & : & \sum_{i} [ q_{L_i} + 3q_{Q_i} ] = 0  \; , 
\\
 \label{eqn:u1y2u1}
\left[U(1)_{Y}\right]^{2} U(1)^{\prime}_{\mbox{\tiny NAF}} & : & 
\sum_{i} [ 2 \times 3 \times \left(\frac{1}{6} \right)^2 q_{Q_i} - 3 \times \left(\frac{2}{3}\right)^2 (-q_{u_i} ) - 3 \times \left( -\frac{1}{3} \right)^2 (-q_{d_i})  \\
& & \qquad \qquad  + 2 \times \left(-\frac{1}{2}\right)^2 q_{L_i} - (-1)^2 (-q_{e_i}) ] = 0 \; , \nonumber 
\end{eqnarray}
\begin{eqnarray}
\label{eqn:u1yu12}
\left[U(1)^{\prime}_{\mbox{\tiny NAF}}\right]^{2} U(1)_{Y} & : & 
\sum_{i} [ 2 \times 3 \times \left( \frac{1}{6} \right) q_{Q_i}^2 - 3 \times \left( \frac{2}{3} \right) \times (-q_{u_i})^2 - 3 \times \left(-\frac{1}{3} \right) (-q_{d_i})^2 \\
& & \qquad \qquad + 2 \times \left(-\frac{1}{2} \right)(q_{L_i})^2 - (-1)(-q_{e_i})^2 ] = 0 \; , 
\nonumber \\
\label{eqn:u1grav}
U(1)^{\prime}_{\mbox{\tiny NAF}}-\mbox{gravity} & : & \sum_{i} [ 6q_{Q_i} + 3q_{u_i} + 3q_{d_i} + 2q_{L_i} + q_{e_i} + q_{N_i}] = 0 \; , 
\\
\label{eqn:u13}
[U(1)^{\prime}_{\mbox{\tiny NAF}}]^{3} & : &  
\sum_{i} [ 3 ( 2 (q_{Q_i})^3 - (-q_{u_i})^3 - (-q_{d_i})^3) + 2(q_{L_i})^3 - (-q_{e_i})^3 - (-q_{N_i})^3] = 0 \; . \;
\end{eqnarray}
Here we follow the standard convention, and all chiral supermultiplets are defined in terms of left-handed Weyl spinors, so that the right-handed singlets are the conjugates of the corresponding SM fields. Therefore, the right-handed fermion singlets carry the opposite $U(1)_{\mbox{\tiny NAF}}^{\prime}$ charges of the corresponding chiral supermultiplets (i.e., $-q_{u_{i}}$, $-q_{d_{i}}$). 

In order to find the solutions to the anomaly cancellation conditions, we find that it is convenient to parametrize the  $U(1)^{\prime}_{\mbox{\tiny NAF}}$ charges in the following way, 
\begin{eqnarray}
q_{Q_1} & = & -\frac{1}{3} q_{L_1} - 2a \; ,  \nonumber \\
q_{Q_2} & = & -\frac{1}{3} q_{L_2} + a + a^{\prime} \; , \nonumber \\
q_{Q_3} & = & -\frac{1}{3} q_{L_3} + a - a^{\prime} \; ,  \nonumber \\
q_{u_1} & = & -\frac{2}{3} q_{L_1} - q_{e_1} - 2b \; ,  \nonumber \\
q_{u_2} & = & -\frac{2}{3} q_{L_2} - q_{e_2} + b + b^{\prime} \; , \nonumber \\
q_{u_3} & = & -\frac{2}{3} q_{L_3} - q_{e_3} + b - b^{\prime} \; , \nonumber  \\
q_{d_1} & = & \frac{4}{3} q_{L_1} + q_{e_1} - 2c \; , \nonumber \\
q_{d_2} & = & \frac{4}{3} q_{L_2} + q_{e_2} + c + c^{\prime} \; , \nonumber  \\
q_{d_3} & = & \frac{4}{3} q_{L_3} + q_{e_3} + c - c^{\prime} \; , \nonumber 
\end{eqnarray}
\begin{eqnarray}
q_{N_1} & = & -2q_{L_1} - q_{e_1} - 2d \; , \nonumber \\
q_{N_2} & = & -2q_{L_2} - q_{e_2} + d + d^{\prime} \; , \nonumber \\
q_{N_3} & = & -2q_{L_3} - q_{e_3} + d - d^{\prime} \; .
\label{eqn:parametz}
\end{eqnarray}
With this parameterization, all anomaly conditions are satisfied except for the $[U(1)^{\prime}_{\mbox{\tiny NAF}}]^{2} U(1)_{Y}$ condition given in Eq. (\ref{eqn:u1yu12}), and the $[U(1)^{\prime}_{\mbox{\tiny NAF}}]^{3}$ condition given in Eq. (\ref{eqn:u13}).

\section{Fermion Mass Hierarchy and Mixing from $U(1)^{\prime}_{\mbox{\tiny NAF}}$ Symmetry}
\label{sec:FermMassMixing}
Given that all three generations of chiral superfields have generation dependent charges under the $U(1)^{\prime}_{\mbox{\tiny NAF}}$ symmetry, the $U(1)^{\prime}_{\mbox{\tiny NAF}}$ symmetry also plays the role of a family symmetry which gives rise to the observed mass hierarchy and mixing angles of the SM fermions. With the experimental constraints on the fermion masses and mixing angles, the number of free parameters in the model is further reduced.
 
In the presence of the $U(1)^{\prime}_{\mbox{\tiny NAF}}$ symmetry, the Yukawa matrices in the superpotential as shown in Eq. (\ref{eqn:sPoten}) are the effective Yukawa couplings generated through higher dimensional operators {\it \`a la} the Froggatt-Nielsen mechanism. As a result, they can be written as powers of the ratio of the flavon fields, $ \Phi $ and $ \Phi^{\prime}$, that breaks the $U(1)^{\prime}_{\mbox{\tiny NAF}}$ symmetry, to the cutoff scale of the $U(1)^{\prime}_{\mbox{\tiny NAF}}$ symmetry, $\Lambda$,
\begin{equation}
\label{eqn:genYukawa1}
Y_{ij} \sim \biggl( y_{ij} \frac{  \Phi }{\Lambda} \biggr)^{3|q_i+q_j+q_H|} \; .
\end{equation} 
Similarly, the $\mu$ term is generated by the higher dimensional operator and it is given by
\begin{equation}
\label{eqn:genmu}
\mu \sim \biggl( \mu_{ud} \frac{  \Phi  }{\Lambda} \biggr)^{3|q_{H_u}+q_{H_d} - 1/3|}  \Phi   \; .
\end{equation} 
The chiral superfield $\Phi$ is a SM gauge singlet whose $U(1)^{\prime}_{\mbox{\tiny NAF}}$ charge is normalized to $-1/3$ in our model. The parameters $y_{ij}$ and $\mu_{ud}$ are coupling constants of order $\mathcal{O}(1)$; $q_i$ and $q_j$  are the $U(1)^{\prime}_{\mbox{\tiny NAF}}$ charges of the chiral superfields of the $i$-th and $j$-th generations of quarks and leptons, and $q_H$ (which can be $q_{H_u}$ or $q_{H_d}$) denotes the $U(1)^{\prime}_{\mbox{\tiny NAF}}$ charges of the up- and down-type Higgses. Note that if $q_{i}+q_{j}+q_{H} < 0$ or $q_{H_{u}} + q_{H_{d}} < 1/3$, then instead of the $\Phi$ field, the field $\Phi^{\prime}$ whose $U(1)^{\prime}_{\mbox{\tiny NAF}}$ charge is $1/3$ is used in Eq. (\ref{eqn:genYukawa1}) or Eq. (\ref{eqn:genmu}), so that the holomorphism of the superpotential is retained. The terms with non-integer $3|q_i+q_j+q_H|$ and $3|q_{H_u}+q_{H_d}|$ are not allowed in the superpotential given that the number of the flavon fields must be an integer. This thus naturally gives rise to texture-zeros in the Yukawa matrices. 

Once the scalar component $\phi$ ($\phi^{\prime}$) of the flavon superfield $\Phi$ ($\Phi^{\prime}$) acquires a vacuum expectation value (VEV), the $U(1)^{\prime}_{\mbox{\tiny NAF}}$ symmetry is broken. Upon the breaking of the $U(1)^{\prime}_{\mbox{\tiny NAF}}$ symmetry and the electroweak symmetry, the effective Yukawa couplings then become,  
\begin{equation}
\label{eqn:genYukLam}
Y_{ij}^{eff} \sim \left(y_{ij}^3 \lambda \right)^{|q_i+q_j+q_H|},
\end{equation}
and the effective $\mu$ term is similarly given by,
\begin{equation}
\label{eqn:genMuLam}
\mu \sim \left(\mu_{ud}^3 \lambda \right)^{|q_{H_u}+q_{H_d}-1/3|} \left< \phi \right> \; ,
\end{equation}
where $\lambda \equiv \left( \left< \phi \right> / \Lambda \right)^3$ or $\lambda \equiv \left( \left< \phi^{\prime} \right> / \Lambda \right)^3$.  
The $U(1)^{\prime}_{\mbox{\tiny NAF}}$ charges thus determine the form of the effective Yukawa matrices: 
For the up-type and down-type quark Yukawa matrices, they are given by 
\begin{eqnarray}
Y_u & \sim & \left(\begin{array}{ccc} \lambda^{|q_{Q_1}+q_{u_1}+q_{H_u}|} & \lambda^{|q_{Q_1}+q_{u_2}+q_{H_u}|} & \lambda^{|q_{Q_1}+q_{u_3}+q_{H_u}|} \\ \lambda^{|q_{Q_2}+q_{u_1}+q_{H_u}|} & \lambda^{|q_{Q_2}+q_{u_2}+q_{H_u}|} & \lambda^{|q_{Q_2}+q_{u_3}+q_{H_u}|} \\ \lambda^{|q_{Q_3}+q_{u_1}+q_{H_u}|} & \lambda^{|q_{Q_3}+q_{u_2}+q_{H_u}|} & \lambda^{|q_{Q_3}+q_{u_3}+q_{H_u}|} \end{array} \right) \; ,
\\
Y_d & \sim & \left(\begin{array}{ccc} \lambda^{|q_{Q_1}+q_{d_1}+q_{H_d}|} & \lambda^{|q_{Q_1}+q_{d_2}+q_{H_d}|} & \lambda^{|q_{Q_1}+q_{d_3}+q_{H_d}|} \\ \lambda^{|q_{Q_2}+q_{d_1}+q_{H_d}|} & \lambda^{|q_{Q_2}+q_{d_2}+q_{H_d}|} & \lambda^{|q_{Q_2}+q_{d_3}+q_{H_d}|} \\ \lambda^{|q_{Q_3}+q_{d_1}+q_{H_d}|} & \lambda^{|q_{Q_3}+q_{d_2}+q_{H_d}|} & \lambda^{|q_{Q_3}+q_{d_3}+q_{H_d}|} \end{array} \right) \; .  
\label{eqn:downYukawa}
\end{eqnarray}
Similarly, the effective charged lepton Yukawa matrix can be written as 
\begin{eqnarray}
\label{eqn:downYukawa}
Y_e & \sim & \left(\begin{array}{ccc} \lambda^{|q_{L_1}+q_{e_1}+q_{H_d}|} & \lambda^{|q_{L_1}+q_{e_2}+q_{H_d}|} & \lambda^{|q_{L_1}+q_{e_3}+q_{H_d}|} \\ \lambda^{|q_{L_2}+q_{e_1}+q_{H_d}|} & \lambda^{|q_{L_2}+q_{e_2}+q_{H_d}|} & \lambda^{|q_{L_2}+q_{e_3}+q_{H_d}|} \\ \lambda^{|q_{L_3}+q_{e_1}+q_{H_d}|} & \lambda^{|q_{L_3}+q_{e_2}+q_{H_d}|} & \lambda^{|q_{L_3}+q_{e_3}+q_{H_d}|} \end{array} \right).  
\end{eqnarray}
The neutrino Dirac and right-handed Majorana Yukawa matrices can be written as
\begin{eqnarray}
\label{eqn:neutDiracYukawa}
Y_{\nu} & \sim & \left(\begin{array}{ccc} \lambda^{|q_{L_1}+q_{N_1}+q_{H_u}|} & \lambda^{|q_{L_1}+q_{N_2}+q_{H_u}|} & \lambda^{|q_{L_1}+q_{N_3}+q_{H_u}|} \\ \lambda^{|q_{L_2}+q_{N_1}+q_{H_u}|} & \lambda^{|q_{L_2}+q_{N_2}+q_{H_u}|} & \lambda^{|q_{L_2}+q_{N_3}+q_{H_u}|} \\ \lambda^{|q_{L_3}+q_{N_1}+q_{H_u}|} & \lambda^{|q_{L_3}+q_{N_2}+q_{H_u}|} & \lambda^{|q_{L_3}+q_{N_3}+q_{H_u}|} \end{array} \right) \; ,
\\
\label{eqn:neutMajRiYukawa}
Y_{N} & \sim & \left(\begin{array}{ccc} \lambda^{|2q_{N_1}+q_{\Psi}|} & \lambda^{|q_{N_1}+q_{N_2}+q_{\Psi}|} & \lambda^{|q_{N_1}+q_{N_3}+q_{\Psi}|} \\ \lambda^{|q_{N_2}+q_{N_1}+q_{\Psi}|} & \lambda^{|2q_{N_2}+q_{\Psi}|} & \lambda^{|q_{N_2}+q_{N_3}+q_{\Psi}|} \\ \lambda^{|q_{N_3}+q_{N_1}+q_{\Psi}|} & \lambda^{|q_{N_3}+q_{N_2}+q_{\Psi}|} & \lambda^{|2q_{N_3}+q_{\Psi}|} \end{array} \right)  \; . 
\end{eqnarray}

Because of the heaviness of the top quark, bottom quark, and tau masses, we assume that 
\begin{equation}
q_{Q_3}+q_{u_3}+q_{H_u} = 0, 
\quad 
q_{Q_3}+q_{d_3}+q_{H_d} = 1, 
\quad
q_{L_3}+q_{e_3}+q_{H_d} = 1,
\end{equation}
leading to no suppression or small suppression in the (3, 3) elements in the corresponding Yukawa matrices. Additionally, to keep the $U(1)^{\prime}_{\mbox{\tiny NAF}}$ symmetry breaking scale high, we choose 
\begin{equation}
q_{L_3}+q_{N_3}+q_{H_u} = 2 \; . 
\end{equation}
To obtain the lepton mass hierarchy and large mixings, we choose the $U(1)^{\prime}_{\mbox{\tiny NAF}}$ charge splittings of the charged leptons so that 
\begin{equation}
q_{L_1} = q_{L_3}+1, \quad 
q_{L_2} = q_{L_3}, \quad 
q_{e_1} = q_{e_3} + 3, \quad 
q_{e_2} = q_{e_3} + 2 \; . 
\end{equation}
The effective charged lepton Yukawa matrix is given by, 
\begin{eqnarray}
\label{eqn:lepYukawaNew}
Y_e & \sim & \left(\begin{array}{ccc} \lambda^{5} & \lambda^{4} & \lambda^{2} \\ \lambda^{4} & \lambda^{3} & \lambda^{1} \\ \lambda^{4} & \lambda^{3} & \lambda^{1} \end{array} \right) \; .
\end{eqnarray}
After putting 8 constraints shown above, we are left with 8 free parameters which are $q_{L_3}$, $q_{e_3}$, $a^{\prime}$, $b$, $b^{\prime}$, $c^{\prime}$, $d$, $d^{\prime}$ and the parametrization equantions ( Eq. (\ref{eqn:parametz})) can be rewritten as
\begin{eqnarray}
q_{Q_1} & = & -\frac{1}{3} q_{L_3} - 2a^{\prime} + 2b - 2b^{\prime} - 2d + 2d^{\prime} + \frac{11}{3} \; ,  \nonumber \\
q_{Q_2} & = & -\frac{1}{3} q_{L_3} + 2a^{\prime} - b + b^{\prime} + d - d^{\prime} - 2\; , \nonumber \\
q_{Q_3} & = & -\frac{1}{3} q_{L_3} - b + b^{\prime} + d - d^{\prime} - 2\; ,  \nonumber \\
q_{u_1} & = & -\frac{2}{3} q_{L_3} - q_{e_3} - 2b - \frac{11}{3} \; ,  \nonumber \\
q_{u_2} & = & -\frac{2}{3} q_{L_3} - q_{e_3} + b + b^{\prime} - 2\; , \nonumber \\
q_{u_3} & = & -\frac{2}{3} q_{L_3} - q_{e_3} + b - b^{\prime} \; , \nonumber \\
q_{d_1} & = & \frac{4}{3} q_{L_3} + q_{e_3} - 2b + 2b^{\prime} - 2c^{\prime} + 2d - 2d^{\prime} + \frac{1}{3} \; , \nonumber \\
q_{d_2} & = & \frac{4}{3} q_{L_3} + q_{e_3} + b - b^{\prime} + 2c^{\prime} - d + d^{\prime} + 4 \; , \nonumber  \\
q_{d_3} & = & \frac{4}{3} q_{L_3} + q_{e_3} + b - b^{\prime} -d + d^{\prime} + 2 \; , \nonumber \\
q_{N_1} & = & -2q_{L_3} - q_{e_3} - 2d -5 \; , \nonumber \\
q_{N_2} & = & -2q_{L_3} - q_{e_3} + d + d^{\prime} - 2 \; , \nonumber \\
q_{N_3} & = & -2q_{L_3} - q_{e_3} + d - d^{\prime} \; .
\label{eqn:newParametz}
\end{eqnarray}
In addition, to generate the neutrino mass hierarchy and mixings, we impose two other requirements,
\begin{equation}
d = -\frac{4}{3} \; , \; \; d^{\prime} =  1 \; ,
\end{equation} 
which lead to 
\begin{equation}
q_{N_{1}} = q_{N_{2}} = q_{N_{3}} = q_{N} \; ,
\end{equation}
and the Dirac mass term for the neutrinos,
\begin{eqnarray}
\label{eqn:neutDiracYukawaNew}
Y_{\nu} & \sim & \left(\begin{array}{ccc} \lambda^{3} & \lambda^{3} & \lambda^{3} \\ \lambda^{2} & \lambda^{2} & \lambda^{2} \\ \lambda^{2} & \lambda^{2} & \lambda^{2} \end{array} \right) \, .
\end{eqnarray}
Furthermore, to make use of the Type-I seesaw mechanism, we assume that the $U(1)^{\prime}_{\mbox{\tiny NAF}}$ charge of the $\Psi$ field is,
\begin{equation}
q_{\Psi} = -4 - 2q_{N} = -4 + 2(2q_{L_3} + q_{e_3} + \frac{7}{3}) \; , 
\end{equation}
such that the neutrino right-handed Majorana mass matrix is allowed. However, this does not reduce the number of the free parameters but gives a democratic RH neutrino Majorana mass matrix,
\begin{eqnarray}
\label{eqn:neutMajRiYukawaNew}
Y_{N} \left< \Psi \right> & \sim & \left(\begin{array}{ccc} \lambda^4 & \lambda^4 & \lambda^4 \\ \lambda^4 & \lambda^4 & \lambda^4 \\ \lambda^4 & \lambda^{4} & \lambda^{4} \end{array} \right) \left< \Psi \right> \; .
\end{eqnarray}
Therefore, the effective light neutrino mass matrix is 
\begin{eqnarray}
\label{eqn:neutMass}
m_{\nu}  \sim  Y_{\nu} Y_{N}^{-1} Y_{\nu}^{T} \frac{v^2}{\left< \Psi \right>}  \sim \left(\begin{array}{ccc} \lambda^2 & \lambda & \lambda \\ \lambda & 1 & 1 \\ \lambda & 1 & 1 \end{array} \right) \frac{v^2}{\left< \Psi \right>} \; .
\end{eqnarray} 
The $U(1)^{\prime}_{\mbox{\tiny NAF}}$ symmetry is broken near the GUT scale ($\left< \Psi \right> \sim 10^{15}$ GeV), and the mass scale of the right-handed neutrino is $\sim 10^{12}$ GeV. Therefore, after the seesaw mechanism takes place, the above mass matrices lead to effective light neutrino masses in the sub-eV range, in addition to a MNS matrix with two large and one small mixing angles.

The $[U(1)^{\prime}_{\mbox{\tiny NAF}}]^{2} U(1)_{Y}$ anomaly cancellation condition, Eq. (\ref{eqn:u1yu12}), is satisfied, if 
\begin{equation}
\label{eqn:bPara}
b = \frac{364 - 114a^{\prime} + 18a^{\prime 2} - 183b^{\prime} + 27a^{\prime}b^{\prime} + 
     18b^{\prime 2} + 96c{\prime} - 27b^{\prime}c^{\prime} + 
     18c^{\prime 2}}{9(-17 + 3a^{\prime} + 6b^{\prime} - 3c^{\prime})} \; .
\end{equation}
The $[U(1)^{\prime}_{\mbox{\tiny NAF}}]^{3}$ anomaly cancellation condition, Eq. (\ref{eqn:u13}), gives rise to a further relation among the parameters, enabling the variable $q_{e_3}$ to be determined in terms of the variables $a^{\prime}$, $b^{\prime}$, $c^{\prime}$ and $q_{L_3}$. These are the only four independent parameters in the model at this stage.

To obtain the observed quark mass hierarchy, we further require 
\begin{equation}
c^{\prime} = -a^{\prime}, \quad b^{\prime} = -1/2 - a^{\prime} \; , \;
\end{equation} 
which further reduce the number of free parameter down to two. Consequently, the effective quark Yukawa matrices can be expressed in terms of a single parameter, $a^{\prime}$.  
Specifically, the effective up-type quark Yukawa matrix is given by,
\begin{eqnarray}
\label{eqn:upYukawaNew}
Y_u & \sim & \left(\begin{array}{ccc} \lambda^{10} & \lambda^{|\frac{7}{2}-\frac{2a^{\prime}}{5}|} & \lambda^{|\frac{13}{2}+\frac{8a^{\prime}}{5}|} \\ \lambda^{|\frac{7}{2}+\frac{2a^{\prime}}{5}|} & \lambda^{|-3|} & \lambda^{|2a^{\prime}|} \\ \lambda^{|\frac{7}{2}-\frac{8a^{\prime}}{5}|} & \lambda^{|-3-2a^{\prime}|} & \lambda^{0} \end{array} \right) \; , \;
\end{eqnarray}
and the effective down-type quark Yukawa matrix is given by,
\begin{eqnarray}
\label{eqn:downYukawaNew}
Y_d & \sim & \left(\begin{array}{ccc} \lambda^{5} & \lambda^{|\frac{19}{2}-\frac{2a^{\prime}}{5}|} & \lambda^{|\frac{15}{2}+\frac{8a^{\prime}}{5}|} \\ \lambda^{|-\frac{3}{2}+\frac{2a^{\prime}}{5}|} & \lambda^{3} & \lambda^{|1+2a^{\prime}|} \\ \lambda^{|-\frac{3}{2}-\frac{8a^{\prime}}{5}|} & \lambda^{|3-2a^{\prime}|} & \lambda^{1} \end{array} \right) \; . \;
\end{eqnarray}
Note that the diagonal elements in $Y_{u}$ and $Y_{d}$ are always allowed, and they give rise to realistic masses for the up-type and down-type quarks. For a wide range of $a^{\prime}$ values, the off diagonal elements of $Y_{u}$ and $Y_{d}$ are forbidden, resulting in a CKM matrix which is proportional to the identity.  To the leading order, this is a good approximation. Non-zero quark mixing may be generated through other effects such as loop contributions.     

In general, with the anomaly cancellation conditions and the aforementioned conditions from realistic fermion masses and mixing, we find a class of  models satisfying all these requirements. These models are specified by two free parameters $a^{\prime}$ and $q_{L_3}$. The corresponding $U(1)^{\prime}_{\mbox{\tiny NAF}}$ charges of the chiral superfields are summarized in Table~\ref{tbl:u1Charge}. 
\begin{table}[b!]
\begin{tabular}{c|c}\hline\hline\
Field & $U(1)^{\prime}_{\mbox{\tiny NAF}}$ charge \\ \hline
$L_1$& $q_{L_1} = 1+q_{L_3}$ \\ \hline 
$L_2$& $q_{L_2} = q_{L_3}$ \\ \hline 
$L_3$& $q_{L_3} = q_{L_3}$ \\ \hline 
$e_1^c$& $q_{e_1} = -(-386375+65664a^{\prime 2}+153000q_{L_3}+1080a^{\prime}(37+48q_{L_3}))/(180(425+144a^{\prime}))$ \\ \hline  
$e_2^c$& $q_{e_2} = -(-309875+65664a^{\prime 2}+153000q_{L_3}+1080a^{\prime}(61+48q_{L_3}))/(180(425+144a^{\prime}))$ \\ \hline
$e_3^c$& $q_{e_3} = -(-156875+65664a^{\prime 2}+153000q_{L_3}+1080a^{\prime}(109+48q_{L_3}))/(180(425+144a^{\prime}))$ \\ \hline
$Q_1$ & $q_{Q_1} = 38/9+2a^{\prime}/5-q_{L_3}/3$ \\ \hline
$Q_2$ & $q_{Q_2} = -41/18+4a^{\prime}/5-q_{L_3}/3$ \\ \hline
$Q_3$ & $q_{Q_3} = (-205-108a^{\prime}-30q_{L_3})/90$\\ \hline
$u_1^c$ & $q_{u_1} = (55296a^{\prime 2}+720a^{\prime}(173+48q_{L_3})+125(-371+816q_{L_3}))/(180(425+144a^{\prime}))$ \\ \hline
$u_2^c$ & $q_{u_2} = (44928a^{\prime 2}+1080a^{\prime}(-69+32q_{L_3})+125(-4349+816q_{L_3}))/(180(425+144a^{\prime}))$\\ \hline   
$u_3^c$ & $q_{u_3} = (96768a^{\prime 2}+720a^{\prime}(217+48q_{L_3})+125(-2513+816q_{L_3}))/(180(425+144a^{\prime}))$\\ \hline   
$d_1^c$ & $q_{d_1} = -(-46625+25344a^{\prime 2}+17000q_{L_3}+480a^{\prime}(107+12q_{L_3}))/(60(425+144a^{\prime}))$\\ \hline
$d_2^c$ & $q_{d_2} = (32275-5760a^{\prime 2}-3400q_{L_3}-72a^{\prime}(63+16q_{L_3}))/(5100+1728a^{\prime})$\\ \hline   
$d_3^c$ & $q_{d_3} = (22075-2304a^{\prime 2}-3400q_{L_3}-96a^{\prime}(-23+12q_{L_3}))/(5100+1728a^{\prime})$\\ \hline  
$\nu_1^{c}$& $q_{N_1} = (-335375+57240a^{\prime}+65664a^{\prime 2})/(180(425+144a^{\prime}))$ \\ \hline
$\nu_2^{c}$& $q_{N_2} = (-335375+57240a^{\prime}+65664a^{\prime 2})/(180(425+144a^{\prime}))$ \\ \hline
$\nu_3^{c}$& $q_{N_3} = (-335375+57240a^{\prime}+65664a^{\prime 2})/(180(425+144a^{\prime}))$ \\ \hline
$H_u$& $q_{H_u} = -(-488375+65664a^{\prime 2}+76500q_{L3}+1080a^{\prime} (5+24q_{L3}))/(180(425+144 a^{\prime}))$ \\ \hline
$H_d$& $q_{H_d} = (65664a^{\prime 2}+1080a^{\prime}(133+24q_{L3})+125 (-643+612q_{L3}))/(180(425+144a^{\prime}))$ \\ \hline
$\Phi$ & $q_{\Phi} = -1/3$\\ \hline 
$\Psi$ & $q_{\Psi} = (182375-109080a^{\prime}-65664a^{\prime 2})/(38250+12960a^{\prime})$\\ \hline \hline
\end{tabular}
\caption{The $U(1)^{\prime}_{\mbox{\tiny NAF}}$ charges of all chiral superfields that are free of all gauge anomalies and give realistic masses and mixing angles for all quarks and leptons, including the RH neutrinos. These charges are parametrized by only two parameters, $a^{\prime}$ and $q_{L_{3}}$.}  
\label{tbl:u1Charge}
\end{table}

\section{Sparticle Mass Spectrum}
\label{sec:SparMass}
One characteristic feature of AMSB in the presence of D-term contributions is the existence of sum rules among squared masses of the sparticles. As the $U(1)_{\mbox{\tiny NAF}}^{\prime}$ symmetry in our model is generation-dependent and non-anomalous, the sum rules in our model are quite distinct from those found in other AMSB models with $U(1)^{\prime}$ symmetry \cite{ref:sumRuleAC}. The anomaly cancellation constraints lead to the D-term contributions among various fields to be cancelled automatically. Hence, the sum of the modified masses squared is still equal to the sum of mass square from the original AMSB contribution. The anomaly cancellation conditions $[SU(3)]^{2} U(1)^{\prime}_{\mbox{\tiny NAF}}$, $[SU(2)_{L}]^{2} U(1)^{\prime}_{\mbox{\tiny NAF}}$, $\left[U(1)_{Y}\right]^{2} U(1)^{\prime}_{\mbox{\tiny NAF}}$, give rise to the following RG invariant mass sum rules, and
\begin{eqnarray}
\label{eqn:sumSu3U1}
\sum_{i=1}^{3} (\bar{m}_{u_i^c}^2 + \bar{m}_{d_i^c}^2 + 2\bar{m}_{Q_i}^2) = \sum_{i=1}^{3} (m_{u_i^c}^2 + m_{d_i^c}^2 + 2m_{Q_i}^2)_{\mbox{\tiny{AMSB}}} \, ,  \\
\label{eqn:sumSu2U1}
\sum_{i=1}^{3} (\bar{m}_{L_i}^2 + 3 \bar{m}_{Q_i}^2) = \sum_{i=1}^{3} (m_{L_i}^2 + 3m_{Q_i}^2)_{\mbox{\tiny{AMSB}}} \, , \\
\label{eqn:sumU1Y2U1}
\sum_{i=1}^{3} (\bar{m}_{u_i^c}^2 + \bar{m}_{e_i^c}^2 - 2\bar{m}_{Q_i}^2) = \sum_{i=1}^{3} (m_{u_i^c}^2 + m_{e_i^c}^2 - 2 m_{Q_i}^2)_{\mbox{\tiny{AMSB}}} 
\, ,
\end{eqnarray}
where terms on the right-handed side are the pure AMSB contributions, which are given in terms of $m_{3/2}^2$ and coefficients that are determined by the low energy dynamics (i.e., the gauge coupling constants and Yukawa coupling constants of MSSM). Similarly, sum rules within each generation can be derived from the $U(1)_{\mbox{\tiny NAF}}^{\prime}$ gauge invariance \cite{ref:sumRuleGI}, 
\begin{eqnarray}
\label{eqn:quhu}
\bar{m}_{Q_i}^2 + \bar{m}_{u_j^c}^2 + \bar{m}_{H_u}^2 & = & (m_{Q_i}^2 + m_{u_j^c}^2 + m_{H_u}^2)_{\mbox{\tiny{AMSB}}} + (q_{Q_i} + q_{u_j} + q_{H_u}) \zeta \; (i, j = 1, 2, 3) \, , \\
\label{eqn:qdhd}
\bar{m}_{Q_i}^2 + \bar{m}_{d_j^c}^2 + \bar{m}_{H_d}^2 & = & (m_{Q_i}^2 + m_{d_j^c}^2 + m_{H_d}^2)_{\mbox{\tiny{AMSB}}} + (q_{Q_i} + q_{d_j} + q_{H_d}) \zeta \; (i, j = 1, 2, 3)  \, , \\
\label{eqn:lehd}
\bar{m}_{L_i}^2 + \bar{m}_{e_j^c}^2 + \bar{m}_{H_d}^2 & = & (m_{L_i}^2 + m_{e_i^c}^2 + m_{H_d}^2)_{\mbox{\tiny{AMSB}}} +(q_{L_i} + q_{e_j} + q_{H_d}) \zeta \; (i, j = 1, 2, 3) \, .
\end{eqnarray}
From Eqs. (\ref{eqn:sumSu3U1}-\ref{eqn:sumU1Y2U1}), we can also derive the sum rules for the physical masses,
\begin{eqnarray}
m_{\tilde{u}_L}^2 + m_{\tilde{u}_R}^2 + m_{\tilde{d}_L}^2 + m_{\tilde{d}_R}^2 + m_{\tilde{c}_L}^2 + m_{\tilde{c}_R}^2 + m_{\tilde{s}_L}^2 + m_{\tilde{s}_R}^2 + m_{\tilde{t}_1}^2 + m_{\tilde{t}_2}^2 + m_{\tilde{b}_1}^2 + m_{\tilde{b}_2}^2   \\
= 2\sum_{i = 1}^{3} (2m_{\tilde{Q}_i}^2 + m_{\tilde{u}_i^c} + m_{\tilde{d}_i^c})_{\mbox{\tiny{AMSB}}} + 2 \sum_{i = 1}^{3} (m_{u_i}^2 + m_{d_i}^2) \, ,\nonumber
\end{eqnarray}
\begin{eqnarray}
m_{\tilde{e}_L}^2 + m_{\tilde{e}_R}^2 + m_{\tilde{\mu}_L}^2 + m_{\tilde{\mu}_R}^2 + m_{\tilde{\tau}_1}^2 + m_{\tilde{\tau}_2}^2 + m_{\tilde{u}_L}^2 + m_{\tilde{u}_R}^2 + m_{\tilde{c}_L}^2 + m_{\tilde{c}_R}^2 + m_{\tilde{t}_1}^2 + m_{\tilde{t}_2}^2  \\
= \sum_{i = 1}^{3} (m_{\tilde{L}_i}^2 + m_{\tilde{e}_i^c}^2 + m_{\tilde{Q}_i}^2 + m_{\tilde{u}_i^c}^2)_{\mbox{\tiny{AMSB}}} + 2\sum_{i = 1}^{3} (m_{e_i}^2 + m_{u_i}^2) \, . \nonumber
\end{eqnarray}

In addition to various sum rules, another characteristic attribute is that the degeneracy of the sfermion masses among the first two generations is lifted. In the generation independent $U(1)^{\prime}$ senario, the first two generations of the sfermions in each sector have the same masses individually. However, in our generation dependent $U(1)^{\prime}$ model, their mass squared splittings are proportional to the $U(1)_{\mbox{\tiny NAF}}^{\prime}$ charge splitting, i.e., $m_{\tilde{f}_2}^2 - m_{\tilde{f}_2}^2 = \zeta (q_{f_2} - q_{f_1})$, which are non-zero. More explicitly, the mass squared splittings are
\begin{eqnarray}
\label{eqn:massSqrSplit}
m_{\tilde{e}_L}^2 - m_{\tilde{\mu}_L}^2 = (q_{L_1} - q_{L_2}) \zeta = \zeta \; , \nonumber \\
m_{\tilde{e}_R}^2 - m_{\tilde{\mu}_R}^2 = (q_{e_1} - q_{e_2}) \zeta = \zeta \; , \nonumber \\
m_{\tilde{u}_L}^2 - m_{\tilde{c}_L}^2 = (q_{Q_1} - q_{Q_2}) \zeta = \left(\frac{13}{2} - \frac{2}{5} a^{\prime}\right) \zeta \; , \nonumber \\
m_{\tilde{u}_R}^2 - m_{\tilde{c}_R}^2 = (q_{u_1} - q_{u_2}) \zeta = \left( \frac{13}{2} + \frac{2}{5} a^{\prime} \right) \zeta \; , \nonumber \\
m_{\tilde{d}_L}^2 - m_{\tilde{s}_L}^2 = (q_{Q_1} - q_{Q_2}) \zeta = \left( \frac{13}{2} - \frac{2}{5} a^{\prime} \right) \zeta \; , \nonumber \\
m_{\tilde{d}_R}^2 - m_{\tilde{s}_R}^2 = (q_{d_1} - q_{d_2}) \zeta =  \left( -\frac{9}{2} + \frac{2}{5} a^{\prime} \right) \zeta \; ,
\end{eqnarray}
and these relations are RG invariant. Therefore, by measuring the mass splittings, we can distinguish various $U(1)_{\mbox{\tiny NAF}}^{\prime}$ models by identifying the charge splittings. 

Here we present a numerical example with $a^{\prime} = -27/5$ and $q_{L_3} = 1/2$, which sloves the slepton mass problem in AMSB by giving rise to positive values to all slepton squared masses. The corresponding $U(1)^{\prime}_{\mbox{\tiny NAF}}$ charges of the chiral superfields are summarized in Table ~\ref{tbl:u1Charge2}. 

With these parameters, only the diagonal terms in the effective up-type and down-type quark Yukawa matrices are allowed, 
\begin{eqnarray}
Y_{u} \sim \mbox{diag}(\lambda^{10}, \lambda^{3}, \lambda^{0}) \; , \\
Y_{d} \sim \mbox{diag}(\lambda^{5}, \lambda^{3}, \lambda) \; ,
\end{eqnarray}
which give rise to the quark mass hierarchy naturally taking into account the $\mathcal{O}(1)$ coefficients. The resulting CKM matrix is an identity, which is a good approximation to the leading order.

\begin{table}[b!]
\begin{tabular}{c|c||c|c}\hline\hline\
Field & $U(1)^{\prime}_{\mbox{\tiny NAF}}$ charge & Field & $U(1)^{\prime}_{\mbox{\tiny NAF}}$ charge \\ \hline
$L_1$& $q_{L_1} = 3/2$ & $Q_1$ & $q_{Q_1} = 853/450$ \\ \hline
$L_2$& $q_{L_2} = 1/2$ & $Q_2$ & $q_{Q_2} = -1522/225$ \\ \hline
$L_3$& $q_{L_3} = 1/2$ & $Q_3$ & $q_{Q_3} = 908/225$\\ \hline
$e_1^c$& $q_{e_1} = 31228381/1586700$ & $u_1^c$ & $q_{u_1} = -21278009/1586700$ \\ \hline
$e_2^c$& $q_{e_2} = 29641681/1586700$ & $u_2^c$ & $q_{u_2} = -28164287/1586700$\\ \hline   
$e_3^c$& $q_{e_3} = 26468281/1586700$ & $u_3^c$ & $q_{u_3} = -40540547/1586700$\\ \hline   
$\nu_1^{c}$& $q_{N_1} = -31757281/1586700$ & $d_1^c$ & $q_{d_1} = 10200251/528900$\\ \hline
$\nu_2^{c}$& $q_{N_2} = -31757281/1586700$ & $d_2^c$ & $q_{d_2} = 548909/21156$\\ \hline   
$\nu_3^{c}$& $q_{N_3} = -31757281/1586700$ & $d_3^c$ & $q_{d_3} = 1390561/105780$\\ \hline  
$H_u$& $q_{H_u} = 34137331/1586700$ & $\Phi$ & $q_{\Phi} = -1/3$\\ \hline 
$H_d$& $q_{H_d} = -25674931/1586700$  & $\Psi$ & $q_{\Psi} = 28583881/793350$\\ \hline \hline
\end{tabular}
\caption{The $U(1)^{\prime}_{\mbox{\tiny NAF}}$ charges of the chiral superfields, corresponding to $a^{\prime} = -27/5$ and $q_{L_{3}} = 1/2$. Note that even though some of the charges for the field $f$ may appear to be vary large $\sim \mathcal{O}(20)$, we have the freedom of choosing an overall gauge coupling constant $g$ to be on the order of $< \mathcal{O}(0.1)$ so that the corresponding gauge coupling of the field $f$, $g_{f} = g \cdot q_{f}$, remains perturbative.}   
\label{tbl:u1Charge2}
\end{table}

Since the $U(1)^{\prime}_{\mbox{\tiny NAF}}$ breaking scale is very high (close to the GUT scale), the $Z^{\prime}$ and the right-handed neutrinos as well as their superpartners are very heavy. As a result, the RGEs below the GUT scale are the same as in the MSSM. Thus with the modification of the scalar masses shown in Eq. (\ref{eqn:FIDMass}) as the boundary conditions at the GUT scale, we obtain the mass spectrum of the sparticles at the SUSY scale utilizing  SoftSUSY 3.1 ~\cite{ref:softSUSY}. Furthermore, we choose $\zeta = 1.5 \times (100 \; \mbox{GeV})^2$, $\tan \beta = 10$ and $\mbox{sign}(\mu) = -1$ and $m_{3/2} = 40$ TeV, without including the CKM mixing in the quark sector. Taking the scalar masses shown in Eq. (\ref{eqn:FIDMass}) as the boundary conditions at the GUT scale, we then run SoftSUSY 3.1 and obtain the sparticle masses at the SUSY breaking scale. The sparticle mass spectrum is summarized in Table~\ref{tbl:mass1}.
\begin{table}[t!]
\begin{tabular}{c|c|c|c|c|c|c|c|c|c|c|c}\hline\hline\
Field & $h_0$ & $H_0$ & $A_0$ & $H^+$ & $\tilde{g}$ & $\chi_1$ & $\chi_2$ & $\chi_3$ & $\chi_4$ & $\chi_1^{\pm}$ & $\chi_2^{\pm}$ \\ \hline
Mass (GeV) & 114.81 & 275.74 & 275.51 & 286.93 & 879.93 & 133.99 & 361.94 & 518.34 & 525.65 & 134.15 & 524.55 \\ \hline \hline 
Field & $\tilde{u}_L$ & $\tilde{u}_R$ & $\tilde{d}_L$ & $\tilde{d}_R$ & $\tilde{c}_L$  & $\tilde{c}_R$ & $\tilde{s}_L$ & $\tilde{s}_R$ & $\tilde{t}_1$ & $\tilde{t}_2$ & $\tilde{b}_1$ \\ \hline
Mass (GeV) & 825.53 & 795.10 & 829.11 & 963.65 & 742.91 & 753.27 & 746.89 & 1014.38 & 366.87 & 780.88 & 745.06 \\ \hline \hline
Field & $\tilde{b}_2$ & $\tilde{e}_L$ & $\tilde{e}_R$ & $\tilde{\mu}_L$ & $\tilde{\mu}_R$ & $\tilde{\tau}_1$ & $\tilde{\tau}_2$ & $\tilde{\nu}_{e_L}$ & $\tilde{\nu}_{{\mu}_L}$ & $\tilde{\nu}_{{\tau}_L}$ & $\Delta m_{\chi_1^{\pm} - \chi_{1}}$\\ \hline
Mass (GeV) & 905.41 & 322.45 & 250.78 & 298.35 & 218.71 & 120.09 & 298.56 & 312.44 & 287.44 & 285.58 & 0.16 \\ \hline \hline
\end{tabular}
\caption{The mass spectrum of the sparticles, with $a^{\prime} = -27/5$, $q_{L_{3}}= 1/2$ and $\zeta = 1.5 \times (100 \; \mbox{GeV})^2$.}  
\label{tbl:mass1}
\end{table}
From the mass spectrum, we observe that the mass splitting between the lightest neutralino and the lightest chargino is very small; it is $\sim 160$MeV. This is consistent with one of the distinguishable properties of AMSB mass spectrum, and it can be used to detect AMSB at the collider experiments. Related collider study can be found in \cite{ref:winoPheno}.

\begin{table}[b]
\begin{tabular}{c|c|c|c|c|c|c}\hline\hline\
$\Delta m^2$&$m_{\tilde{e}_L}^2 - m_{\tilde{\mu}_L}^2$& $m_{\tilde{e}_R}^2 - m_{\tilde{\mu}_R}^2$ & $m_{\tilde{u}_L}^2 - m_{\tilde{c}_L}^2$ & $m_{\tilde{d}_L}^2 - m_{\tilde{s}_L}^2$ & $m_{\tilde{u}_R}^2 - m_{\tilde{c}_R}^2$ & $m_{\tilde{d}_R}^2 - m_{\tilde{s}_R}^2$ \\ \hline
$\times (100GeV)^2$&$1.496$& $1.506$ & $1.296$ & $1.296$ & $6.477$ & $-10.035$ \\ \hline \hline 
\end{tabular}
\caption{The mass squared differences between the first two generations of sparticles.}  
\label{tbl:massSplit}
\end{table}
In addition, we have shown numerically that the mass squared differences between the first two generations agree with the mass squared splittings predicted in Eqs. (\ref{eqn:massSqrSplit}). This is shown in the Table \ref{tbl:massSplit} for the specific set of  parameters chosen above.

In the numerical example presented above, stau is the lightest supersymmetric particle (LSP). For this scenario to be viable, R-parity must be broken. There also exists parameter space in our model which predicts neutralino being the LSP and thus R-parity can be retained. This is achieved, for example, by having $\zeta = 1.7 \times (100 \; \mbox{GeV})^2$ while keeping all other parameters the same. The corresponding sparticle mass spectrum is given in Table \ref{tbl:mass2}.
\begin{table}[t!]
\begin{tabular}{c|c|c|c|c|c|c|c|c|c|c|c}\hline\hline\
Field & $h_0$ & $H_0$ & $A_0$ & $H^+$ & $\tilde{g}$ & $\chi_1$ & $\chi_2$ & $\chi_3$ & $\chi_4$ & $\chi_1^{\pm}$ & $\chi_2^{\pm}$  \\ \hline
Mass (GeV) & 114.22 & 163.05 & 162.28 & 180.81 & 879.85 & 133.71 & 360.71 & 488.91 & 497.51 & 133.86 & 495.61  \\ \hline \hline 
Field & $\tilde{u}_L$ & $\tilde{u}_R$ & $\tilde{d}_L$ & $\tilde{d}_R$ & $\tilde{c}_L$ & $\tilde{c}_R$ & $\tilde{s}_L$ & $\tilde{s}_R$ & $\tilde{t}_1$ & $\tilde{t}_2$ & $\tilde{b}_1$ \\ \hline
Mass (GeV) & 825.20 & 790.01 & 828.77 & 978.97 & 730.85 & 742.13 & 734.89 & 1035.46 & 321.22 & 781.79 & 747.97  \\ \hline \hline
Field & $\tilde{b}_2$ & $\tilde{e}_L$ & $\tilde{e}_R$ & $\tilde{\mu}_L$ & $\tilde{\mu}_R$ & $\tilde{\tau}_1$ & $\tilde{\tau}_2$ & $\tilde{\nu}_{e_L}$ & $\tilde{\nu}_{{\mu}_L}$ & $\tilde{\nu}_{{\tau}_L}$ & $\Delta m_{\chi_1^{\pm} - \chi_{1}}$\\ \hline
Mass (GeV) & 914.58 & 347.57 & 273.19 & 322.26 & 239.96 & 142.89 & 322.04 & 338.27 & 312.13 & 310.42 & 0.15 \\ \hline \hline
\end{tabular}
\caption{The mass spectrum of the sparticles, with $a^{\prime} = -27/5$, $q_{L_{3}}= 1/2$ and $\zeta = 1.7 \times (100 \; \mbox{GeV})^2$.}  
\label{tbl:mass2}
\end{table}

\section{Conclusion}
\label{sec:Conclusion}
We propose a MSSM model expanded by a non-universal, non-anomalous $U(1)^{\prime}_{\mbox{\tiny NAF}}$ symmetry. All anomaly cancellation conditions are satisfied with no exotics other than the three right-handed neutrinos. The $U(1)^{\prime}_{\mbox{\tiny NAF}}$ symmetry plays the role of the family symmetry, giving rise to realistic masses and mixing angles for all SM fermions. Furthermore, the FI-D terms associated with the $U(1)^{\prime}_{\mbox{\tiny NAF}}$ symmetry give rise to additional contributions to the slepton masses, rendering them all positive. In a RG invariant way, this thus solves the slepton mass problem in AMSB models. The anomaly cancellation conditions give rise to very stringent constraints on the $U(1)^{\prime}_{\mbox{\tiny NAF}}$ charges of the chiral superfields. We found charges that satisfy all anomaly cancellation conditions and fermion mass and mixing angles, and at the same time solving the slepton mass problem.  While these rational charges are rather complicated, mainly because of the $[U(1)^{\prime}_{\mbox{\tiny NAF}}]^{3}$ anomaly cancellation condition, the differences among the charges are quite simple. The $U(1)^{\prime}_{\mbox{\tiny NAF}}$ charges also dictate the mass spectrum of the sparticles. 

\begin{acknowledgments}
We thank David Sanford, Nick Setzer, and Yuri Shirman for useful discussions. 
The work was supported, in part, by the National Science Foundation under Grant No. PHY-0709742,  PHY-0970173, and 1066293, as well as the hospitality of the Aspen Center for Physics. 
\end{acknowledgments}

\end{document}